# User Requirements for Software Game Process: An Empirical Investigation

Saiqa Aleem[1,*], Luiz Fernando Capretz[2], Faheem Ahmed[3], Shuib Basri[4]

[1]Zayed University, Abu Dhabi, 144534, United Arab Emirates
`Saiqa.Aleem@zu.ac.ae`
[2]Western University, London ON N6G5B9, Canada
`lcapretz@uwo.ca`
[3]Thompson Rivers University, Kamloops BC V2C 0C8, Canada
`fahmed@tru.ca`
[4]Universiti Teknologi PETRONAS, Malaysia
`shuib_basri@utp.edu.my`

**Abstract.** This study attempts to provide a better understanding of the user dimension as a factor in software game success. It focuses mainly on an empirical investigation of the effect of user factors on the software game development process and finally on the quality of the resulting game. A quantitative survey was developed and conducted to identify key user dimensions. For this study, a survey was used to test the research model and hypotheses. The main contribution of this paper is to investigate empirically the influence of user key factors on software game development process that ultimately results in a higher quality final product. The results provide evidence that game development organizations must deal with multiple user key factors to remain competitive and handle high pressure in the software game industry.

## 1    Introduction

### 1.1    A Subsection Sample

In today's world, software game users are seeking more realistic and interactive elements in software games. Nevertheless, the current game development process is unable to accommodate this requirement. Exploring diverse user preferences for software games provides a significant benefit for the game development process by generating valuable insights. The Software Game Development Industry (SDGI) produces highly interactive software such as video games, online games, and mobile games for a wide variety of platforms, including consoles, PCs, mobile devices, and Web browsers. Furthermore, the current Software Game Development (SGD) process consists mainly of a synopsis, background research, script writing, visualization and concept art, level and interaction design, animation, programming, media editing, integration, testing, and publishing.

In the past, researchers highlighted the concept of a user-centered approach to the SGD process. Many attempts have been made to propose methods to capture user perspectives, but very few considered the importance of user preferences during the game development process. The following five important factors were identified from the literature as elements that can directly or indirectly contribute to the development of good-quality software games from a user perspective. Most of the related work done in the past was limited in scope [1] [2], and validation of the proposed models is still an open question.

Therefore, no research has been done to date on including user-centered factors in the software game development process. Consequently, it is necessary to assess key success factors from a user perspective and to search for a new approach that will include user needs in the SGD process. This study will identify key factors empirically from the user perspective, an effort that will ultimately help improve the quality of the SGD process. To identify key factors, a quantitative survey was conducted, and the results are reported in this work. The survey was used to test the research model and several hypotheses. Finally, the results show that consideration of key factors from a user perspective is an important game development choice.

This study will capture the important factors about user's preferences that what they expect to have in a game. Ultimately, identified factors will help developers to consider them while developing games and finally, contributing to the development of high-quality software games.

## 2      Hypothesis

The main objective of the proposed research model is to analyse the associations and interrelationships among the important factors of software games from a user perspective and their influence on software game performance in the SDGI market. The concepts found in the game development literature and in studies, addressing the user perspective on software games provided the theoretical foundation for the proposed research model. Most considered one or two user factors in software game success or performance and examined their impact. We selected user engagement as a dependent variable in our research model; the key user factors are: Game Engagement, Game Enjoyment, Gamer Characteristics, Ease of Use, Socialization:. Hence, Hypothesis 1 can be stated as follows:

**Hypothesis 1:** User engagement factor is important for the success of a software game in the SDGI. In order to identify that users of software game consider engagement an important factor for their favourite game.

**Hypothesis II:** User engagement has a positive and significant effect on software game success. Therefore, to find out that enjoyment is an important factor for any user of software game, we select it as another independent variable in software game success.

**Hypothesis III:** The characteristics of a game are important for the success of the software game. Based on a literature review of the importance of game characteristics, this study has considered game characteristics as another independent variable that is considered important by its user and involved in software game success.

**Hypothesis IV:** Ease of use has a positive and significant impact on the success of software games. To find out that ease of use factor in a software game impact the purchasing behaviour or not, we considered ease of use as an independent variable for this study.

**Hypothesis V:** Social interaction is an important attribute for software game success. Social interaction feature has also been considered as another independent variable involved in software game success, to know that is it considered mandatory by its users or not.

## 3 Research Model

This is the first study that highlights key factors in the software game development process from a user perspective and that examines their influence on software game performance in the market. This study has empirically investigated the associations and influences of key factors from user perspective and software game performance in the SDGI. The theoretical research model used for empirical investigation evaluates the association of various independent variables emerging from game development and the user satisfaction literature on the dependent variable, the success of the final product, i.e., the software game, in the SDGI. Overall, the goal of this study is to investigate and address the following research question:

**Research Question:** What are the key user factors that influence their buying decision and provide motivation to play software games, and what is their impact on overall software game success in the SDGI?

The research model comprises of a total of five independent variables: game engagement, game enjoyment, game characteristics, ease of use, and socialization, and one dependent variable: software game success. The following multiple linear regression equation represents the research model and is given below as Eq. 1:

$$\text{Software game success} = \alpha_0 + \alpha_1 f_1 + \alpha_2 f_2 + \alpha_3 f_3 + \alpha_4 f_4 + \alpha_5 f_5, \quad (1)$$

where $\alpha_0, \alpha_1, \alpha_2, \alpha_3, \alpha_4, \alpha_5$ are coefficients and $f_1$-$f_5$ are the five independent variables.

## 4 Research Methodology

The development and growth of software games have been phenomenal. Nowadays, every home has one or more gamers. The targeted respondents were players or users of software games. Initially, the authors started blogs in software game development communities for data collection. A structured survey questionnaire was also developed. The questionnaire was placed on various blogs on the internet, and some personalized emails were sent to software game groups on social media such as LinkedIn and Facebook. After two and one-half months of investigation, 469 responses had been collected, of which 389 (82.94%) were valid. Participants completed the overall questionnaire under the agreement that their identities would be kept confidential.

First, the authors analysed the basic information from the collected data related to each player's region, the game platform(s) used by the player, and the game genre(s) liked by the player. The respondents came from Asia, Africa, Europe, South America, Australia, and North America, as displayed in Fig. 1. The participants in the survey played games on different platforms such as kiosks and standalone devices, the Web, social networks, consoles, PC/Macs, and mobile phones. Fig. 2 shows the different platforms used and the percentage of respondents using each. The game genres played by the respondents included puzzles, action or adventure, racing, sports, music-based, strategy/role-playing, and other categories. The total percentage of respondents playing each game genre is depicted in Fig. 3. The participants considered for this study were those who played games at least weekly.

### 4.1 Measuring Instrument

Data were collected for this study to analyze the key user factors and the perceived level of game success as identified in the research model taking into consideration the key factors: Game Engagement, Game Enjoyment, Game Characteristics, Ease of Use, and Socialization. The data collection instrument developed for this study was a structured questionnaire and is presented in the Appendix.

Users who play games at least on a weekly basis were considered for this study and were first asked to what extent they considered the identified key factors they considered important for their preferred software game. Second, they were asked about game performance and its relation to success in the SDGI. The questionnaire as developed used the five-point Likert scale, and for each statement, the respondents were required to state their level of agreement or disagreement. Twenty-two items were developed to evaluate the independent variables (the user key factors), and five items were used for the dependent variable (game performance).

The comprehensive list of measuring items for each factor was derived after a detailed review of the literature related to user factors for software games. The five-point Likert scale used for the questionnaire ranged from (1) meaning "strongly disagree" to (5) meaning "strongly agree" and was linked with each item. The measuring items in the questionnaire were numbered sequentially from 1 to 22. The dependent variable, game success, was measured for user's enjoyment, attention, game attributes, ease of use, and socialization based on a multi-item five-point Likert scale. The measuring items for the dependent variable were numbered sequentially and separately from one to five. All items were specifically written for this study and are presented in the Appendix.

### 4.2 Reliability and Validity Analysis

Quantitative research is performed to present reliable and valid research findings. To obtain such findings, empirical studies include two essential measures of precision: reliability and validity. Reliability means that a test or questionnaire is considered reliable if the obtained results are consistent for the test or questionnaire upon re-administration or repetition. It is a measure of consistency. Validity refers to the degree to which a test or questionnaire measures what it claims to measure. It is a measure of agreement between the true value and the measured value. For this empirical study, analyses of both reliability and validity were done for the specifically designed measuring instrument and based on the most commonly used approaches.

In order to carry out an internal consistency analysis, the reliability of multi-scale measurement items for the five user factors was evaluated using Cronbach's alpha coefficient [3]. First, the sample dataset was evaluated using Cronbach's alpha; if any assessment item affected the Cronbach's alpha of its related category, it was excluded from the measuring instrument. All the assessment items were found reliable in the sample test. After this, the whole dataset was evaluated. The reliability of the five user factors was reflected by Cronbach's alpha ranging from 0.65 to 0.80 as given in Table 1.

Many researchers have reported criteria for satisfactory values of Cronbach's alpha based on their findings. Nunnally and Bernste [4] suggested that a satisfactory value for the reliability coefficient must be 0.70 or higher for any measuring instrument. Some researchers have recommended that a reliability co-efficient of 0.55 or higher is satisfactory [5]), and few agreed on to cut-off at 0.60 [6]. Hence, all the variable items developed for the measuring instrument could be considered reliable. To analyze validity, a principal component analysis [7] was conducted for the five user factors, with results reported also in Table 1.

Campbell and Fiske [8] considered that if the scale items were highly correlated and if in a given assembly, they moved in the same direction then convergent validity had occurred. To determine the construct validity of the PCA-based measuring instrument, the eigenvalues [9] were used as a reference point. For empirical investigation in this study, the Kaiser Criterion [10] was used. The Kaiser Criterion aims to keep any value greater than one for any component. Eigenvalue analysis indicated that of the five variables, four formed a single factor, whereas the game characteristics loaded onto two factors, both with eigenvalues greater than one. The reported convergent validity of this study is therefore considered as adequate.

**Table 1.** Cronbach's alpha coefficient and principal component analysis of five variables.

| User factors | Item no. | Coefficient α | PC eigenvalue |
|---|---|---|---|
| Game engagement | 1–5 | 0.71 | 1.49 |
| Game enjoyment | 6–9 | 0.85 | 1.57 |
| Game characteristics | 10–15 | 0.74 | 1.01 |
| Ease of use | 16–19 | 0.70 | 1.16 |
| Socialization | 20–21 | 0.65 | 1.61 |

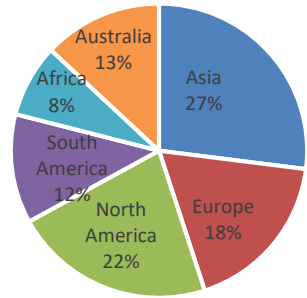

**Fig. 1.** Percentages of respondents by region.

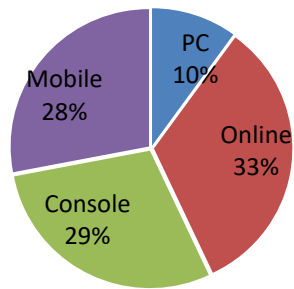

**Fig. 2.** Percentages of respondents for each game platform used.

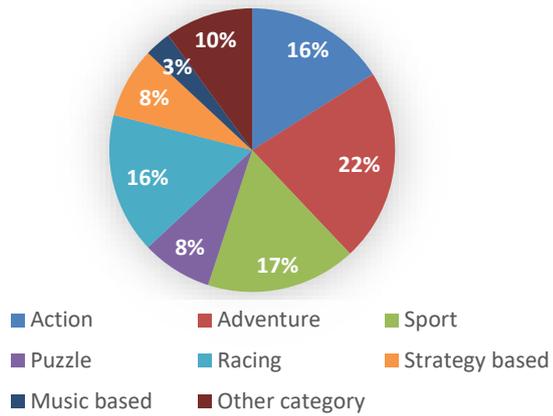

**Fig. 3.** Percentages of respondents based on the game genres they play.

**4.3 Data Analysis Techniques**

For this empirical research, various statistical methods were used to carry out data analysis. Initially, the data analysis was divided into three phases to assess the significance of hypotheses H1–H5. Parametric statistics and normal distribution tests were conducted in phase I. Partial least squares (PLS) analysis was used in phase II as a non-parametric statistical approach. Both approaches were used to address the question of external validity threats. In the measuring instrument, for each dependent and independent variable, multiple items were used. A composite value was obtained by aggregating respondents' ratings for each dependent and independent variable. Pearson's correlation is a statistical approach that measure the linear relationship between two continuous random variables. It does not undertake data normality although it does assume finite covariance and variance. On the other hand, spearman's correlation describes the monotonic relationship between two continuous random variables and also robust to outliers. The number of respondents are sufficient to carryout data analysis but as compared to the area of research, this much number of respondents still considered as small sample size.

In order to address this consideration, we employed both parametric and non-parametric statistical approaches. The Pearson correlation coefficient with the one-tailed t-test was evaluated for each hypothesis H1–H5 in phase I. In phase II, the Spearman correlation coefficient (a non-parametric statistic) was used to test hypotheses H1–H5. Phase III consisted of PLS-based techniques for testing the research model developed from H1–H5. The PLS technique was used in phase III to increase the reliability of the results. It was also reported by Fornell and Bookstein [11] and Joreskog and Wold [12] that PLS methods were helpful in cases of non-normal distribution, complexity, small sample size, and low-theoretical-information datasets. For statistical calculations, the Minitab 17 software was used.

**5   Data Analysis and Results**

Four phases were considered in the investigation.

**5.1   Phase I of Hypothesis Testing**

The Pearson correlation coefficient (a parametric statistical approach) was used to examine and test hypotheses H1–H5. It was examined between the key factors from the user's perspective (independent variables) and software game success (dependent variable) in the research model. The significance level based on the p-value was used to accept or reject each hypothesis. A p-value of less than 0.05 was considered good enough to accept the hypothesis; in case of a value greater than 0.05, the hypothesis would be rejected [13]. Table 2 shows the hypothesis testing results for the Pearson correlation coefficient.

Hypothesis H1 was accepted because the Pearson correlation coefficient for game engagement and game success was positive (0.65) at $p<0.05$. For hypothesis H2, concerning game enjoyment and game success, the Pearson correlation coefficient was also positive (0.75) at p<0.05, and therefore hypothesis H2 was also accepted. Hypothesis H3, regarding game characteristics and game success, was accepted due to its smaller p-value (0.04). Hypothesis H4, concerning ease of use and game success, was also accepted based on its positive Pearson correlation coefficient (0.79) at p<0.05. Hypothesis H5, concerning socialization and game success, was rejected based on its p-value greater than 0.05 (0.29). In short, hypotheses H1, H2, H3, and H4 were accepted and found to be statistically significant. Hypothesis H5 was not supported statistically and was therefore rejected.

### 5.2 Phase II of Hypothesis Testing

The Spearman correlation coefficient (a nonparametric statistical approach) was used in phase II to test hypotheses H1–H5. The results for the Spearman correlation coefficient are also reported in Table 2. The Spearman correlation coefficient for hypothesis H1 was found statistically significant (0.65) at p<0.05, and therefore H1 was accepted. The Spearman correlation coefficient for game enjoyment and game success (hypothesis H2) was also found to be positive (0.74) at p<0.05 and statistically significant. The relationship between game characteristics and game success (hypothesis H3) was found statistically significant due to its Spearman correlation coefficient (0.56) at p<0.05 and was accepted. For hypothesis H4, the Spearman correlation coefficient was positive at p<0.05, and therefore H4 was accepted. Hypothesis H5 concerning socialization and game success was rejected due to its low coefficient (0.30) at p>0.05. Hence, in short, hypotheses H1, H2, H3, and H4 were accepted and found statistically significant. Hypothesis H5 was not supported statistically and was rejected.

**Table 2.** Hypothesis testing using parametric and nonparametric correlation coefficients.

| Hypothesis | Key factor | Pearson coefficient | Spearman coefficient |
|---|---|---|---|
| H1 | Game engagement | 0.65* | 0.65* |
| H2 | Game enjoyment | 0.75* | 0.74* |
| H3 | Game characteristics | 0.55* | 0.56* |
| H4 | Ease of use | 0.79* | 0.78* |
| H5 | Socialization | 0.29** | 0.30** |

*Significant at $p<0.05$
**Insignificant at $p>0.05$

### 5.3 Phase III of Hypothesis Testing

The partial least squares (PLS) approach was used in phase III to test the hypothesis. It was used to overcome the limitations associated with parametric and nonparametric statistical approaches used in phases I and II and also to cross-validate the results of phases I and II. For direction and significance analysis of hypotheses H1–H5, PLS was used. The dependent variable (game success) was selected as the response variable and the individual user factors designated as the predicate variable. The structural test results of phase III, including the observed values of R2, the path coefficient, and the F-ratio, are presented in Table 3.

Game engagement (H1) had a positive path coefficient of 0.65, an R2 = 0.34, and an F-ratio = 18.10 and was found statistically significant at p<0.05. The path coefficient value for game enjoyment (H2) was observed to be 0.95, R2 was 0.29, and the F-ratio was 18.51, and therefore H2 was also found significant at p<0.05. The game characteristics hypothesis (H3) had a positive path coefficient of 1.16, a low R2 of 0.37, and an F-ratio of 35.52 and was judged significant because the p-value was less than 0.05. Ease of use (H4) had a path coefficient of 0.94, an R2 of 0.62, and an F-ratio of 100.38 and was found significant at p<0.05. Socialization (H5) had a path coefficient of 0.02, an R2 = 0.37, and an F-ratio = 35.52 and was therefore found insignificant at $p>0.05$.

### 5.4 Research Model Testing

The linear regression for the research model is presented in Eq. 1. The proposed research model was evaluated to provide empirical evidence that user factors are substantially important and must be considered to develop good-quality software games. The test procedure analyzed the model coefficient values, the regression analysis, and the direction of the associations. The dependent variable (game success) was designated as the response variable and the other independent variables (all the key user factors) as predicate variables. The results of the regression analysis model are reported in Table 4.

The path coefficients of four of the five variables (game engagement, game enjoyment, game characteristics, ease of use) were positive and were found statistically significant at p<0.05. The path coefficient of socialization was positive, but was found not statistically significant at p<0.05 which made this factor insignificant in the research model. The overall R2 value of the research model was 0.76, and the adjusted R2 value was 0.78 with an F-ratio of 22.36, which was significant at $p<0.05$.

Table 3. PLS regression results for hypothesis testing.

| Hypothesis | Factors | Path coefficient | $R^2$ | $F$-Ratio |
|---|---|---|---|---|
| H1 | Game engagement | 0.65 | 0.34 | 18.10* |
| H2 | Game enjoyment | 0.95 | 0.29 | 18.51* |
| H3 | Game characteristics | 1.16 | 0.37 | 35.52* |
| H4 | Ease of use | 0.94 | 0.62 | 100.38* |
| H5 | Socialization | 0.02 | 0.01 | 0.01** |

*Significant at $p<0.05$
**Insignificant at $p>0.05$

Table 4. Linear regression analysis of the research model.

| Model coefficient name | Model coefficient | Coefficient value | $t$-value |
|---|---|---|---|
| Game engagement | $\alpha_1$ | 0.33 | 1.76* |
| Game enjoyment | $\alpha_2$ | 0.78 | 3.98* |
| Game characteristics | $\alpha_3$ | 0.70 | 3.65* |
| Ease of use | $\alpha_4$ | 0.20 | 0.72* |
| Socialization | $\alpha_5$ | 0.16 | 1.09** |
| Constant | $\alpha_0$ | 0.45 | 1.09* |
| $R^2$ | | 0.76 | Adjusted $R^2$ 0.78 |
| $F$-ratio | | 22.36* | |

*Significant at $p<0.05$
**Insignificant at $p>0.05$

## 6  Discussions

This research is a step towards the understanding as it will help software organizations understand the relationships and interdependences between key factors from user's perspective and software game success in the market. This research is the first empirical investigation of user factors in relation to game success and provides an opportunity to explore associations between them empirically. The observed results support the theoretical assertions made here and provide the very first evidence that the consideration of key user factors while developing games are important for the success of any software game. This could well result in institutionalizing the software game development approach, which in turn has a high potential to maximize profits.

The game enjoyment factor has been found to have a positive impact on software game success. Users of software games are motivated to play games because they want to experience enjoyment, and the literature has shown that enjoyment is a positive reaction of a player during a game play session. The enjoyment factor is important to consider in the game development process. Game developers have mainly been using usability guidance tools as heuristics to develop games that have an enjoyment factor. However, what developers think about the enjoyment factor in a game may not necessarily match the user's expectation. It is important to consider the user perspective on the software game enjoyment factor and analyze what users think is important about this factor.

In developing software games, the game characteristics are very important and can influence the user's experience, as has been shown in the literature. This hypothesis was also supported by empirical investigation. Research into game characteristics also must take user perspective into consideration and analyze how much weight users assign to these characteristics. These issues are often discussed by game players and designers, but are seldom written about in any formal way. By emphasizing these basic player-centric concepts, this study may help persuade developers that they may find solutions to design problems by looking at what game users feel about game characteristics.

The ease-of-use factor plays a significant role in the software game development process. In software games, ease of use consists of all attributes of the software game that help its user to control and operate the game easily, within or outside the gameplay. Ease of use is the most talked about and least understood aspect of software games. In this empirical investigation, a positive association was found between the ease-of-use factor and software game success. The literature has also shown that software game users prefer to use and purchase software games that are easy to use (MAC or PC and/or games on a console such as their iPad, mobile phone, Sony Playstation, Microsoft Xbox, or Nintendo Wii).

It is commonly assumed that the social interaction attribute of software games is crucial to their success. It is also assumed that it is helpful to document the gaming experience with the participation of virtual communities, either for MMPROG or for single-player games. Very few researchers have studied the socialization attribute of software games. Because of all the assumptions used, it is important to study the user perspective on the socialization attributes of software games and to identify what users think about these attributes for software games. The findings of this empirical investigation do not support a statistically positive relationship between socialization and software game success. The direction of the association was positive, but the required statistical level of confidence was not supported. For these reasons, the hypothesis that socialization factors are considered very important by users for the success of software games was statistically rejected.

Overall, the findings of the study are important for the development of good quality software game. Rapid and continual changes in technology and intense competition not only affect the business, but also have a great impact on development activities. To deal with this strong competition and high pressure, game development organizations must continuously assess their activities and adopt a proper evaluation methodology. Use of a proper assessment methodology will help the organization identify its strengths and weaknesses and provide guidance for improvement. However, the fragmented nature of the game development process requires a comprehensive evaluation strategy which has not yet been entirely explored. The findings of the study will help the game development organizations to look for contributing key success factors from the user perspective. This study is a part of a larger project aiming to propose software game maturity assessment model [14]. The user perspective is one of the identified dimensions out of developer [15], business [16] and the process itself. The findings of this study also provide the justification to include these factors in the process of assessment methodology.

### 6.1 Limitations and Threats to External Validity

For software engineering processes or product investigations, various empirical approaches are used, such as case studies, metrics, surveys, and experiments. However, certain limitations are associated with empirical studies and with this study as well. Easterbrook et al. [17] suggested four criteria for the validity of empirical studies: internal validity, construct validity, external validity, and reliability. Wohlin et al. [18] stated that the generalization of experimental results to industrial practice by researchers is mostly limited by threats to external validity. For this study, measures were taken to address external threats to validity. The random sampling method was used to select respondents from all around the world. Open-ended questions were also included in the questionnaire.

The choice and selection of independent variables were one of the limitations of this study. To analyze the association and impact of factors for software game success, five independent variables were included. However, other key factors may exist which have a positive association with and impact on software game success, but due to the presence of the selected five variables in the literature, they were included in the study. In addition, other key factors may exist, such as regionally or environmentally based choices, which may have a positive impact on software game success, but were not considered in this study. Furthermore, the focus of this study was only on user factors for software game success.

In software engineering, the increased popularity of empirical methodologies has raised concerns about ethics. However, this study has adhered to all the applicable ethical principles to ensure that it would not violate any experimental ethics guidelines. Regardless of its limitations, this study has contributed to the SGD process and has helped game development organizations understand the user dimension of software games.

## 7      Conclusions

Game development is a complex task, and measuring the user experience of games poses an additional challenge. For the successful development of good-quality software games, game developers must consider and explore all related dimensions as well as discussing with all the stakeholders involved. This study provides a better understanding of the user dimension of software game success and explores the impact of key factors on the success of software games from a user perspective. This study has mainly tried to answer the research question that was posed earlier in this paper and to analyze the impact of user key factors on game success.

The results of this empirical investigation have demonstrated that user key factors are very important and play a key role in the success of any software game. Moreover, the results showed that game engagement, game enjoyment, game characteristics, and ease of use are positively associated with software game success in the SDGI market. The empirical investigation found no strong association or impact between socialization and software game success.

In the field of game development, this research is the first of its kind and will help game developers and game development organizations to achieve a better understanding of the user's perspective on software games. To improve the current game development process and develop good-quality games, it is important for developers to consider the user preferences as well as others. This study has provided the empirical evidence and justification to include factors from the user perspective in evaluating the user dimension of game development process maturity.

## Appendix

Evaluation of key user success factors identified through literature review. The questionnaire objective is to find out which factors have a positive impact on software game success. Please select the correct scale based on your best knowledge.

| Key factors for game development process from user perspective (Likert Scale) | | | | | | |
|---|---|---|---|---|---|---|
| (1 = strongly disagree; 2= disagree; 3 = neutral; 4= agree; 5 strongly agree) | 1 | 2 | 3 | 4 | 5 | n. a. |
| Game Engagement | | | | | | |
| 1 | I like to play games that provide lots of stimuli (motivations) from different sources. | | | | | |
| 2 | I believe that a game must have all the tasks of equal importance. | | | | | |

| | | | | | | | |
|---|---|---|---|---|---|---|---|
| 3 | The game workload is according to my cognitive (skills or abilities), perceptual (being transported into the real world), and memory limits. | | | | | | |
| 4 | I must be consciously aware of my acts in the virtual world. | | | | | | |
| 5 | I prefer to play games in which I feel full absorption in the game play. | | | | | | |
| Game Enjoyment | | | | | | | |
| 6 | I believe that game play must be interesting and attractive. | | | | | | |
| 7 | I like to play games that maintain my curiosity about all the levels of games. | | | | | | |
| 8 | A good game must provide information about a player's performance and positive competence. | | | | | | |
| 9 | Games must provide encouragement to each player. | | | | | | |
| Game Characteristics | | | | | | | |
| 10 | The game must have challenges that match the skills of its players. | | | | | | |
| 11 | I like to learn a game in a fun and enjoyable way. | | | | | | |
| 12 | I like to be rewarded according to the efforts and skills I have developed. | | | | | | |
| 13 | Game goals should be provided early and in a clear way as I progress in the game. | | | | | | |
| 14 | I prefer to play games that have a good storyline. | | | | | | |
| 15 | I prefer games that provide a feeling of full control and independence. | | | | | | |
| Ease of use | | | | | | | |
| 16 | A good game should provide tutorial support. | | | | | | |
| 17 | Control consistency in terms of internal and external navigational support for menus should be included. | | | | | | |
| 18 | I like to play games that have hints and goal support. | | | | | | |
| 19 | I prefer games with subtitles and magnifier support. | | | | | | |
| Socialization | | | | | | | |
| 20 | I prefer games that support co-operation and competition between players. | | | | | | |
| 21 | I like to participate in online social communities for software games. | | | | | | |
| 22 | I like to play games that support interaction between players. | | | | | | |
| Digital Game Success | | | | | | | |
| 23 | The software game kept the player's attention and focus all the time. | | | | | | |
| 24 | The software game provided enjoyment that helped the player to feel deep and effortless involvement. | | | | | | |
| 25 | The good characteristics of games in terms of challenges, feedback, clear goals, interactive interface features, and bug reporting attracted players. | | | | | | |
| 26 | The software game provided easy-to-use gameplay so that players felt a sense of control over their actions within the game. | | | | | | |
| 27 | The software game provided socialization attributes that created and supported opportunities for social interaction. | | | | | | |